\documentclass[aps,epsfigure,twocolumn,superscriptaddress,showkeys,nofootinbib]{revtex4-1}
\usepackage[colorlinks=true,linkcolor=blue,urlcolor=blue,citecolor=blue,pdfusetitle]{hyperref}
\usepackage[utf8]{inputenc}
\usepackage[english]{babel}
\usepackage{amsmath}
\usepackage[caption = false]{subfig}
\usepackage{comment}
\usepackage{graphicx, float, epstopdf}
\usepackage{blindtext}
\usepackage[table,xcdraw]{xcolor}
\usepackage{amsfonts}
\usepackage{bbm}
\usepackage{amssymb}
\usepackage{enumerate}
\usepackage{color}
\usepackage{latexsym}
\usepackage{physics}
\usepackage{times,txfonts}
\usepackage[normalem]{ulem}
\graphicspath{ {./figures/} }

\usepackage{amsmath}
\usepackage{booktabs}

\usepackage{tikz}
\usetikzlibrary{quantikz}

\usepackage[toc]{appendix}

\begin{document}

\title{Exploring Quantum Neural Networks for Demand Forecasting}

\author{Gleydson Fernandes de Jesus}
\email{gleydson.jesus@fieb.org.br}
\affiliation{QuIIN - Quantum Industrial Innovation, EMBRAPII CIMATEC Competence Center in Quantum Tecnologies, SENAI CIMATEC, Av. Orlando Gomes, Salvador, BA 1845, Brazil.}
\affiliation{Latin America Quantum Computing Center, SENAI CIMATEC, Salvador, Bahia, Brazil.}

\author{Maria Heloísa Fraga da Silva}
\affiliation{QuIIN - Quantum Industrial Innovation, EMBRAPII CIMATEC Competence Center in Quantum Tecnologies, SENAI CIMATEC, Av. Orlando Gomes, Salvador, BA 1845, Brazil.}
\affiliation{Latin America Quantum Computing Center, SENAI CIMATEC, Salvador, Bahia, Brazil.}
\affiliation{Universidade Federal do Oeste da Bahia, Barreiras, Bahia, Brazil.}

\author{Otto Menegasso Pires}
\affiliation{QuIIN - Quantum Industrial Innovation, EMBRAPII CIMATEC Competence Center in Quantum Tecnologies, SENAI CIMATEC, Av. Orlando Gomes, Salvador, BA 1845, Brazil.}
\affiliation{Latin America Quantum Computing Center, SENAI CIMATEC, Salvador, Bahia, Brazil.}

\author{Lucas Cruz da Silva}
\affiliation{Robotics Department, SENAI CIMATEC, Salvador, Bahia, Brazil.}

\author{Clebson dos Santos Cruz}
\affiliation{Universidade Federal do Oeste da Bahia, Barreiras, Bahia, Brazil.}

\author{Valéria Loureiro da Silva}
\affiliation{QuIIN - Quantum Industrial Innovation, EMBRAPII CIMATEC Competence Center in Quantum Tecnologies, SENAI CIMATEC, Av. Orlando Gomes, Salvador, BA 1845, Brazil.}

\begin{abstract}

Forecasting demand for assets and services can be addressed in various markets, providing a competitive advantage when the predictive models used demonstrate high accuracy. However, the training of machine learning models incurs high computational costs, which may limit the training of prediction models based on available computational capacity. In this context, this paper presents an approach for training demand prediction models using quantum neural networks. For this purpose, a quantum neural network was used to forecast demand for vehicle financing. A classical recurrent neural network was used to compare the results, and they show a similar predictive capacity between the classical and quantum models, with the advantage of using a lower number of training parameters and also converging in fewer steps. Utilizing quantum computing techniques offers a promising solution to overcome the limitations of traditional machine learning approaches in training predictive models for complex market dynamics.
\end{abstract}

\keywords{Demand forecasting, Quantum machine learning, Quantum finance.}

\maketitle

\section{Introduction}

A common problem experienced by companies is the financial market uncertainty \cite{Alessandri2019Financial, Ghosal2019The}, which makes accurate forecasting and budgeting challenging, becoming a risk to investments and financial stability \cite{Alessandri2019Financial, Ghosal2019The, Stockhammer2010Financial, Kumar2020Stock}. Companies must adapt quickly to these changes in order to remain competitive and mitigate any negative impacts \cite{Guo2023Research, Zedginidze2023Strategic}. In this context, demand forecasting is defined as a predictive analysis strategy used to overcome this challenge using traditional computational methods or more advanced technologies, including machine learning \cite{Spiliotis2020Comparison, Abbasimehr2020An, Aktepe2021Demand}.

In this scenario, the demand estimation process helps companies internally plan to meet market needs \cite{Subramanian2020Demand, Aktepe2021Demand}. It involves forecasting the number of services or products a company will sell in a future period \cite{Ferreira2016Analytics}. The duration of this period can be customized and may vary based on the company's size and objectives \cite{Dorrington2020Beyond}.

In order to help managers make more assertive decisions about team planning and demand management, forecasting takes into account internal and external factors that meet customer needs \cite{abolghasemi2020demand}. The benefits encompass enhancements in efficiency, operational performance, and the supply chain as it predicts the number of goods to be sold and, subsequently, the amount that has to be manufactured \cite{aamer2020data}, thereby preventing the occurrence of insufficient or excessive production. 

Furthermore, demand forecasting can be influenced by either a qualitative or quantitative methodology \cite{kerkkanen2010improving}. The first case is a {superficial subjective analysis} of customer behavior and market trends. In the second case, statistical data is compared and analyzed from both the sales history and customer base to provide a more in-depth picture of the future \cite{Spiliotis2020Comparison}. {Accordingly, demand forecasting was traditionally based on statistical methods and expert opinion, which often made it difficult to capture complex patterns and dynamic market trends \cite{jeyaraman2024machine}. 
Given this scenario, the implementation of machine learning algorithms in demand forecasting has resulted in notable improvements in forecast accuracy \cite{jeyaraman2024machine}.}

In this context, classical machine learning (ML) is a widely used computational tool for solving the problem of demand forecasting \cite{Feizabadi2020Machine}. It helps identify patterns in large volumes of historical data and make accurate forecasts. However, as data volumes increase and models become more complex, the processing limitations of classical models become more apparent due to the difficulty of capturing multiple characteristics of high dimensional data \cite{Oner2020Combining}. Conversely, quantum computing has emerged as a promising solution. The ability of quantum computers to process information in parallel offers a significant advantage, allowing for the efficient and quick analysis of massive datasets and optimization of machine learning models \cite{Kendon2020Quantum, cerezo2022challenges}. 

Thus, \textit{quantum machine learning} (QML) emerges as a promising alternative that can accelerate information processing and provide notable improvements in the machine learning research area ~\cite{2022Houssein, 2022Cerezo}. Recent advancements in this area suggest that the integration of quantum computing with machine learning is poised to lead to groundbreaking developments in technology and data analysis \cite{Melnikov2023Quantum, Ciliberto2017Quantum}. This approach offers a way to address the difficulties presented by conventional machine learning techniques \cite{Ciliberto2017Quantum,PhysRevA.103.032430}, such as increased learning duration caused by the expanding amount of data \cite{Caro2022},  for instance.  Hence, quantum computing and quantum machine learning (QML) have recently experienced increased utilization across various domains, including finance~\cite{2023Cherrat}. 

In this regard, this work presents an application for QML to predict vehicle financing demand using a quantum neural network. For this purpose, we utilized a dataset obtained from the Brazilian bank BV, containing financing data and other relevant features collected from 2019 to 2023. This data was pre-processed, and smaller sets were extracted using feature reduction techniques \cite{Shafizadeh-Moghadam2021Fully, Ma2019Dimension}. The quantum neural network was trained on this pre-processed data to accurately predict vehicle financing demand, showcasing the potential of quantum computing in enhancing predictive analytics. The results obtained from this study demonstrate the promising capabilities of QML models in solving complex real-world problems such as financial forecasting. The integration of quantum computing in predictive analytics can revolutionize the way financial institutions make decisions and manage risks. By leveraging the power of QML, banks can gain a competitive edge in the market by making more accurate and timely predictions. Therefore, this research highlights the importance of leveraging quantum computing in the financial sector to improve decision-making processes, which represents a significant advancement in the field of predictive analytics.

\section{Quantum Neural Networks}
\label{sec-qml}

Quantum neural networks (QNNs) represent a new approach to machine learning, combining classical data processing with the power of quantum computing~\cite{2022Avramouli, 2022Cerezo, 2022Houssein}. Despite their classical foundations, QNNs are considered \textit{pure quantum} models because their execution depends on classical computing only for circuit preparation and statistical analysis~\cite{2023Combarro}. These QNNs fall under Variational Quantum Algorithms, employing Parameterized Quantum Circuits (PQCs) known as \textit{ans\"{a}tze} (plural of ansatz), which are trained using classical optimization techniques. The behavior of quantum neural networks (QNNs) reflects that of classical neural networks, consisting of three main stages: data preparation, data processing, and data output~\cite{2023Combarro}.

In the data preparation stage, the classical input is encoded into a quantum state using a feature map, a circuit parameterized exclusively by the original data~\cite{schuld2021machine}. This coding facilitates the integration of the classical information into the quantum structure of the QNN. In particular, classical data may require pre-processing, such as normalization or scaling, to optimize the coding process \cite{schuld2021machine, 2023Ogur}.

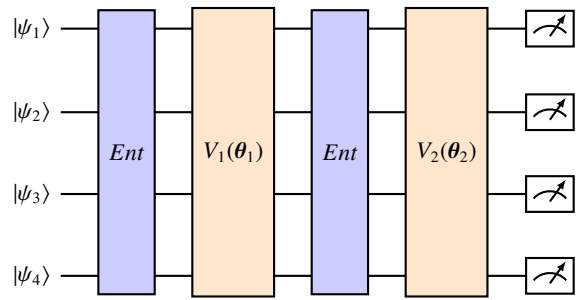
\begin{figure}[ht]
    \centering
    \begin{quantikz}
        \lstick{$\ket{\psi_1}$} & \gate[4, style={fill=blue!20}]{Ent} & \gate[4, style={fill=orange!20}]{V_1(\boldsymbol{\theta}_1)} & \gate[4, style={fill=blue!20}]{Ent} & \gate[4, style={fill=orange!20}]{V_2(\boldsymbol{\theta}_2)} & \meter{} \\
        \lstick{$\ket{\psi_2}$} &  &  &  &  & \meter{} \\
        \lstick{$\ket{\psi_3}$} &  &  &  &  & \meter{} \\
        \lstick{$\ket{\psi_4}$} &  &  &  &  & \meter{}
    \end{quantikz}
    \caption{A two-layered ansatz applied to four qubits. Each layer is defined by a variational circuit $V_j$ dependent on some parameters $\boldsymbol{\theta}_j$. The circuits $Ent$ are used to entangle the qubits, and the state $\ket{\psi}^{\otimes n}$ denotes the output of the feature map.}
    \label{fig:qml}
\end{figure}

Subsequently, in the data processing stage, the QNN operates within the framework of its ansatz. Usually structured as a layered variational circuit, the ansatz consists of multiple layers, each defined by an independent parameter vector. Variational circuits $V_j$ dependent on these parameters make up each layer, with layers of entanglement $Ent$ interspersed. An example of a quantum ansatz is shown in Figure~\ref{fig:qml}. The ansatz effectively processes quantum-coded data, taking advantage of entanglement and parameterized gates for computation.

Finally, in the data output stage, the processed quantum state is converted into a classical output via a final layer~\cite{2023Combarro}. This operation is adapted to the specific problem being addressed. For example, in a binary classification, the expected value of a single qubit selected in the measurement can be used as the output \cite{schuld2021machine}. Overall, QNNs offer a promising path for quantum-assisted machine learning, uniting classical and quantum paradigms to address complex computational tasks \cite{schuld2021machine, 2022Cerezo, 2022Houssein, 2023Ogur}.

{In this article, we use a QNN, whose architecture is described in Section \ref{subsec-QNNarchitecture}, to perform the task of forecasting the demand for used vehicle financing in Brazil.}

\section{Quantum Data Analysis and Model Implementation}
\label{Methodology}

\subsection{Data Scaling and Selection Techniques}

The case analyzed in this paper was the forecasting of used car prices in Brazil from May 2022 to April 2023. The data used for training covered the period from January 2019 to April 2022 and it was provided by the Brazilian bank BV, so the application is of practical interest in the financial sector. In all, 25 features were provided for the training. However, 6 features were discarded because they did not contain data for 2019. The remaining 19 features were subjected to a feature reduction process using the Principal Component Analysis (PCA) 
so that the features with the greatest variance were selected. PCA is a statistical technique that enables the reduction of data dimensionality while preserving variance. It identifies principal components, linear combinations of initial features, ranked based on their total variance, ensuring significant new features are identified \cite{Shafizadeh-Moghadam2021Fully}.

\begin{figure}[H]
    \centering
    \includegraphics[scale = 0.5]{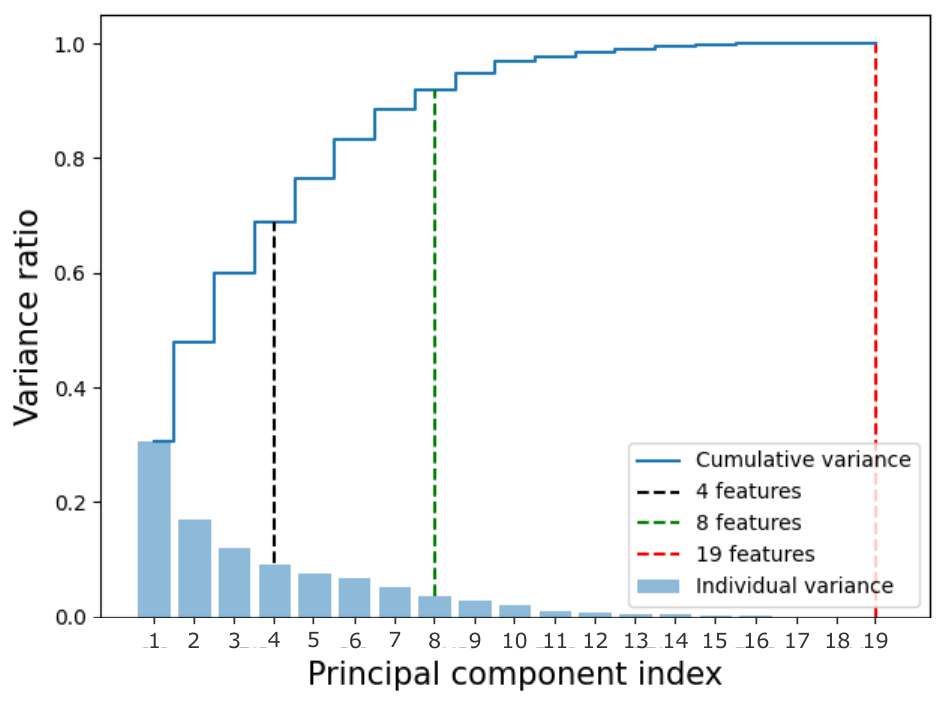}
    \caption{Cumulative variance of the data. The x-axis represents the component index, while the y-axis represents the variance. The sets used in this article were marked in black (4 features), green (8 features), and red (19 features, the complete dataset). The bars represent the individual variance of each component, while the blue line represents the cumulative variance.}
    \label{fig:variancia}
\end{figure}

The smallest sets used had 4 and 8 features, which represented 68.69\% and 92.03\% of the total variance of the dataset, respectively. In addition, the complete dataset represents 100\% of the total variance of the distribution. The cumulative variances are shown in figure \ref{fig:variancia}. 

After reducing the number of features, the data was standardized according to the expression \cite{2023Ogur}:

\begin{equation}
    \hat{x} = \frac{x - \mu}{\sigma},
\end{equation} 

where $x$ is the original data, $\mu$ is the mean of the values, $\sigma$ the standard deviation and $\hat{x}$ the standardized data. This standardization assumes that the distribution of the data is approximately normal.
Standardizing the data helps to ensure that all variables are on the same scale, which is important for many machine learning algorithms since it reduces the scale of the data set, thereby decreasing the differences in scale between features and avoiding biases in features with larger scales. This process makes it easier to compare and interpret the coefficients of different features in the model. 

\subsection{Quantum Neural Network Architecture}
\label{subsec-QNNarchitecture}

In quantum neural network architecture, qubits are used to represent data and parameters in the model \cite{Cong2018Quantum}. By leveraging quantum superposition and entanglement, quantum neural networks have the potential to outperform classical neural networks in certain tasks by processing information in a more efficient way \cite{2023Combarro, Narayanan2000Quantum}. This architecture holds promise for solving complex problems in fields such as optimization, machine learning, and cryptography. 
The Quantum Neural Network model used to process this data is presented in the figure \ref{fig:rede_usada}. In this model, the feature mapping is done through $RY$ rotation gates, performed after initializing the circuits in uniform superposition through $Hadamard$ gates. Two variational layers were considered and are represented in the figure \ref{fig:camadas_emaranhamento}.

\begin{figure}[H]
\begin{center}
\begin{quantikz}
& \gate{H}\gategroup[4,steps=2,style={dashed,rounded corners,fill=red!20, inner xsep=2pt},background,label style={label position=below,anchor=north,yshift=-0.2cm}]{{\sc Feature Map}} & \gate{R_y(x_0)} & \gate[4]{Entanglement}\gategroup[4,steps=2,style={dashed,rounded corners,fill=blue!20, inner xsep=2pt},background,label style={label position=below,anchor=north,yshift=-0.2cm}]{{\sc Ansatz}} & \gate{R(\theta_0, \phi_0, \omega_0)} & \meter{}\gategroup[4,steps=1,style={dashed,rounded corners,fill=green!20, inner xsep=2pt},background,label style={label position=below,anchor=north,yshift=-0.2cm}]{{\sc Measure}} & \\
& \gate{H} & \gate{R_y(x_1)} &  & \gate{R(\theta_1, \phi_1, \omega_1)} & \meter{} & \\
& \gate{H} & \gate{R_y(x_2)} &  & \gate{R(\theta_2, \phi_2, \omega_2)} & \meter{} & \\
& \gate{H} & \gate{R_y(x_3)} &  & \gate{R(\theta_3, \phi_3, \omega_3)} & \meter{} & 
\end{quantikz}
\end{center}
\caption{{Variational Quantum Circuit. Hadamard gates layer 
prepares the qubits in uniform superposition, Ry gates (red) encode the data in qubits, variational layer, or ansatz (blue) entangle the qubits and applies parametrized rotations, {where $\theta_i$, $\phi_i$ and $\omega_i$ represent, respectively, the rotation angles in the x, y, and z axes in each qubit $i$, and are the trainable parameters of the model}. Measurement layer (green) collapse the qubits, generating the outputs \cite{2023Ogur}.}}
\label{fig:rede_usada}
\end{figure}
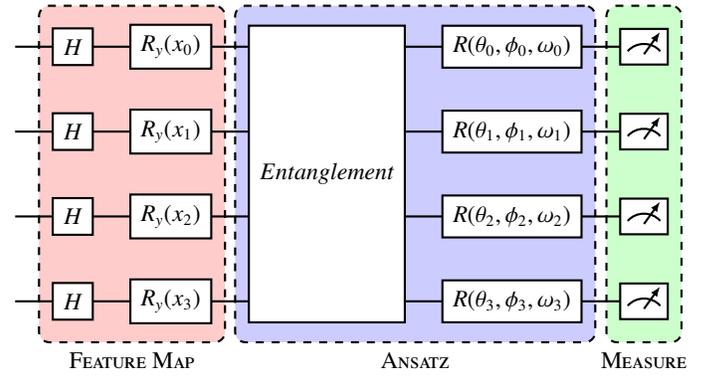

\begin{figure}[H]
\begin{center}
\label{emaranhamento}
\begin{quantikz}
& \ctrl{1}\gategroup[4,steps=2,style={dashed,rounded corners,fill=gray!20, inner xsep=2pt},background,label style={label position=above,anchor=north,yshift=+0.4cm}]{{(a)}} & \qw & \qw & \\
& \targ{} & \ctrl{1} & \qw & \\
& \ctrl{1} & \targ{} & \qw & \\
& \targ{} & \qw & \qw &
\end{quantikz}
\hfill
\begin{quantikz}
& \ctrl{1}\gategroup[4,steps=5,style={dashed,rounded corners,fill=gray!20, inner xsep=2pt},background,label style={label position=above,anchor=north,yshift=+0.4cm}]{{(b)}} & \qw & \qw & \targ{} & \qw & \qw & \\
& \targ{}  & \ctrl{1} &    \qw   &  \qw      &  \qw  & \qw  & \\
&   \qw    & \targ{}  & \ctrl{1} &  \qw      &  \qw  & \qw  & \\
&   \qw    &   \qw    & \targ{}  &  \ctrl{-3} &   \qw  & \qw  &  
\end{quantikz}
\label{fig:camadas_emaranhamento}
\caption{{Entanglement Layers used in variational circuit (Figure \ref{fig:rede_usada}). In (a), here named "entanglement layer 1", the qubits are entangled in pairs, and these pairs are subsequently tied together. In (b), here named "entanglement layer 2", the qubits are entangled in a cascade. Adapted from \cite{2023Ogur}.}}
\end{center}
\end{figure}
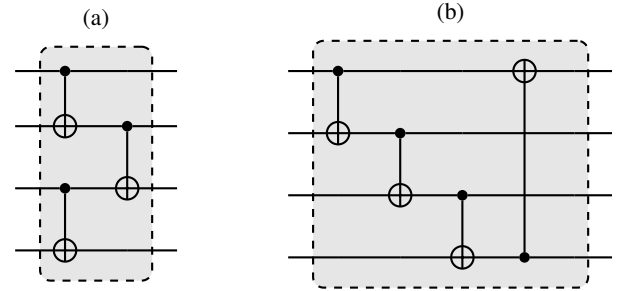 

In order to take into account the real monetary losses due to inaccurate results, the models were analyzed based on the mean absolute error obtained in each experiment. In addition, the accuracy of the results obtained through the heuristics used was analyzed by calculating the standard deviation obtained during the 12 months of testing. Each experiment was run 10 times.

\section{Dataset and Preprocessing}
\label{sec-dataset}

In addition to the variable to be predicted, the dataset made available for the research contained 25 economic features relevant to the proposed model, with 52 samples collected monthly from January 2019 to April 2023. The model's prediction variable is the daily average of used vehicle financing each month and was also part of the dataset made available.

During the period covered in the database, the world faced the COVID-19 pandemic, and governments around the world shut down their countries' economies to promote social isolation. For this reason, we can consider this period to be anomalous. 

However, removing this data would mean reducing the already scarce number of samples available for training, so it has been kept in its original form. In addition, as it is understood that not the pandemic itself but its impacts on economic indicators (contained in the database used) are the most relevant features for the predictive model in question, no new features were added to the database that could bring new information about the pandemic scenario. It should also be considered that the proposed model must be robust to any factors that impact the financial market and the inability to predict such occurrences makes it necessary to measure them indirectly through their influence on economic indicators. 

When analyzing the data made available for the research, it was noticed that some features did not contain data for 2019. As the number of instances for training was already low, it was necessary to remove these features since the alternatives to such exclusion would be to exclude the data for the entire year 2019 or to infer the missing data, which could make the model biased and was therefore not done. After excluding the 6 features that did not contain data for 2019, a feature reduction was carried out using the PCA method \cite{Shafizadeh-Moghadam2021Fully, Ma2019Dimension}. 

In addition, the PCA method allows for the reduction of the number of features in a dataset by applying a transformation to the coordinate axes. This transformation generates new axes that point in the directions of the greatest variance in the datasets. The directions of greatest variance are the main components of the models since they supposedly have more information to extract during the training process and can then be used in this process. The PCA method was used through the scikit-learn \cite{scikit-learn} machine learning library.

Based on the data available and after eliminating the data that did not have values for 2019, 3 different sets of data were generated, with 4, 8, and 19 features, which represent 68.89\%, 92.03\%, and 100\% of the system's total variance, respectively. These sets were generated to assess the impact of adding new features to the models. 

\section{Results}
\label{sec-results}

The results obtained from the quantum experiments are presented in Section \ref{Quantum experiments}, while those from the classical experiments are detailed in Section \ref{Classical experiments}. The number of variational layers selected for the quantum experiments includes configurations of 1, 3, and 5 layers, chosen to assess their impact on performance.

The measurement results are presented in terms of the distributions observed. Training and test errors are illustrated graphically, focusing on the average daily financing obtained. Additionally, the mean monthly absolute errors are provided in tables, offering a comprehensive view of the variations and trends.

\subsection{Quantum experiments} \label{Quantum experiments}

\subsubsection{4 Features}

In the first considered case, the dataset was reduced from 19 initial features to 4 features. Similarly to the other quantum experiments, we considered the two quantum networks presented in section \ref{Methodology}, as well as a classical RNN. 

\begin{widetext}

\begin{figure}[H]
    \centering
    \includegraphics[scale = 0.5]{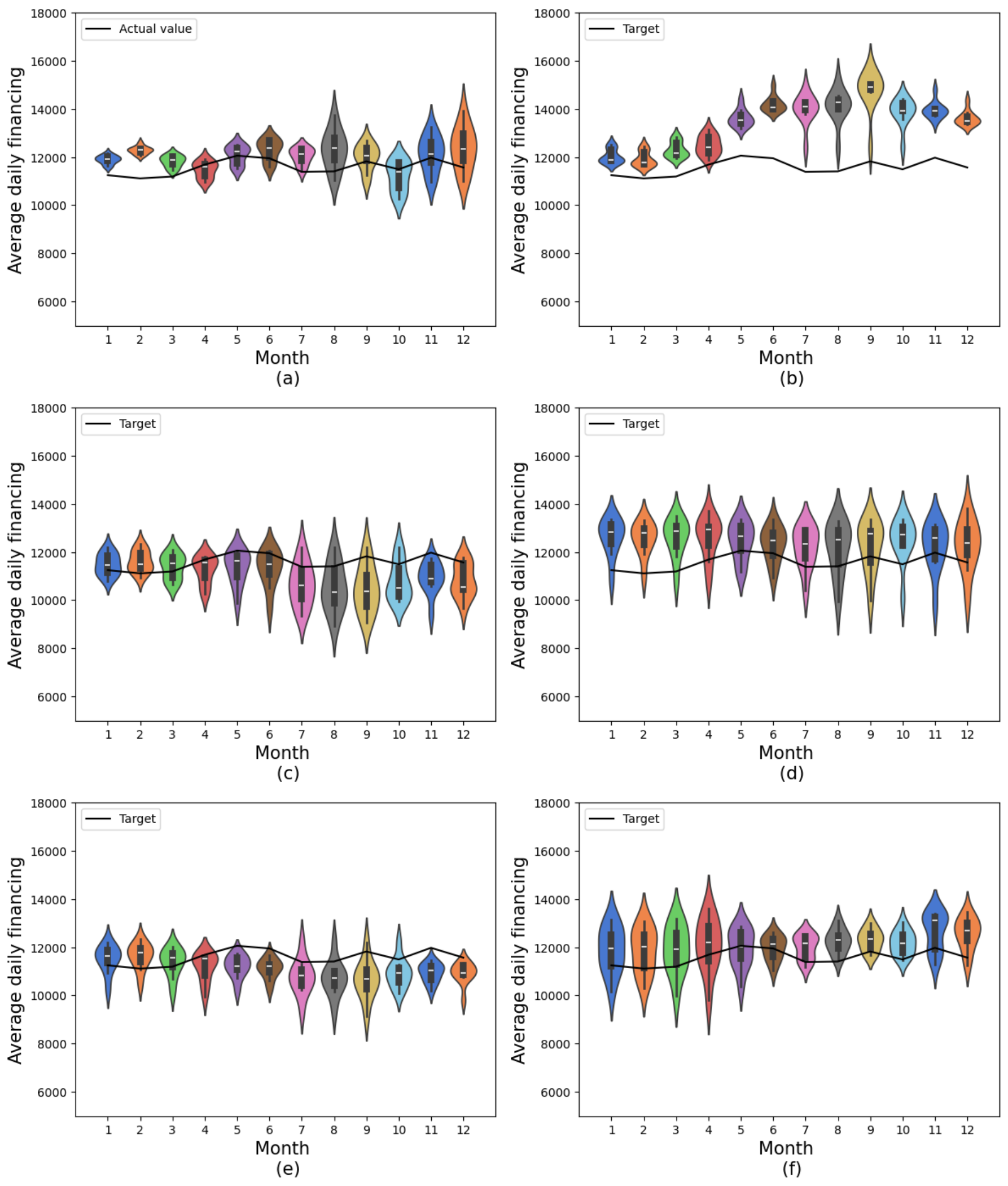}
    \caption{Predictions for quantum models with 4 features. Figure (a) shows the quantum experiment 1 with 1 layer, figure (b) quantum experiment 2 with 1 layer, figure (c) quantum experiment 1 with 3 layers, figure (d) quantum experiment 2 with 3 layers, figure (e) quantum experiment 1 with 5 layers and (f) quantum experiment 2 with 5 layers. The x-axis shows the model's training months, while the y-axis represents average daily financing. The distributions obtained from 10 experiments are shown in the colored violin graphs, while the actual values are shown in the black line.}
    \label{fig:4featuresresults}
\end{figure}   

\end{widetext}

Figure \ref{fig:4featuresresults} shows the results of the quantum networks obtained in the two experiments using the 4-feature dataset and varying the number of variational layers. The actual values to which the predictions should approximate are shown in the black curves. The simulation environment is discussed in the section \ref{environment}, and the convergence of the models is discussed in the section \ref{Convergence}. The standard deviation for each month is presented in table \ref{tab:std4quantum}, in appendix \ref{subsection:quantumstd}, and the monthly mean absolute error for the two experiments are presented in table \ref{tab:mae4quantum}, in appendix \ref{subsection:quantummae}. The cumulative variance of the data contained in this database concerning the original database with nineteen features is 68.69\%.

Violin graphs were used to present the results of the predictions. These graphs show the density distributions through their contours. In this way, wider points in the figures represent a greater density of data, while sparser and more distant points represent outliers. In addition, these graphs can reveal multimodal trends in the distributions when there is more than one widening point. The black boxplot in the center of the figures shows the median of the distributions through a white line in the boxes, and the first and third quartiles are represented respectively through the lower and upper edges of the box plot.

Considering realistic scenarios where errors are inevitable, it may be preferable to err upwards or downwards, depending on the market and the agents involved. Nonetheless, it is important to consider that errors above the target may be a warning of the need for greater production of a certain product or availability of services, while errors below the target may represent missed opportunities to sell products or services with a higher demand than predicted by the machine learning models.

Among the quantum models consisting of four features, experiment 1, involving a single variational layer, generally showed the best accuracy, with a minimum mean absolute error of 299.90 and a 12-month mean absolute error of 682.52 ± 284.14.

In the second experiment, the best result was the one in which 5 variational layers were considered, where the lowest mean absolute error obtained was 411.97, and the monthly mean was 785.98 ± 192.21. In addition, this model shows considerably less variation than the results obtained with just one variational layer, as well as a considerably lower mean. 

Furthermore, in both cases, the means of the predictions were above the actual values, which could imply that there is a greater supply of used car finance than there is actual demand if these indicators are the only ones considered.

\subsubsection{8 Features}

In the second considered case, the number of features in the database was reduced to eight. This value was chosen as an intermediate value between the four features used initially and the final nineteen features, given that the cumulative variance for eight features was 92.03\%.

\begin{widetext}

\begin{figure}[H]
    \centering
    \includegraphics[scale = 0.5]{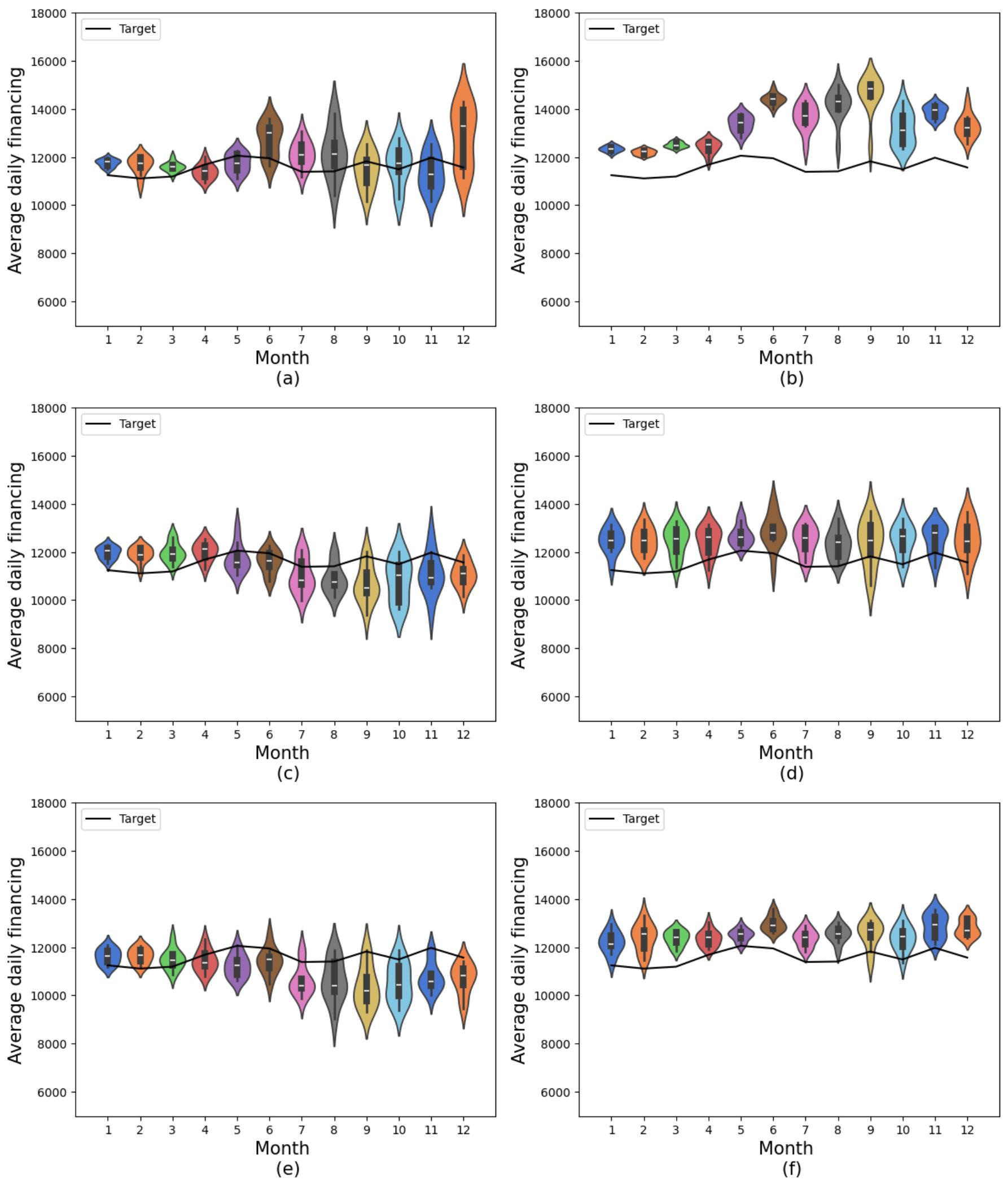}
    \caption{Predictions for quantum models with 8 features. Figure (a) shows the quantum experiment 1 with 1 layer, figure (b) quantum experiment 2 with 1 layer, figure (c) quantum experiment 1 with 3 layers, figure (d) quantum experiment 2 with 3 layers, figure (e) quantum experiment 1 with 5 layers and (f) quantum experiment 2 with 5 layers. The x-axis shows the model's training months, while the y-axis represents average daily financing. The distributions obtained from 10 experiments are shown in the colored violin graphs, while the actual values are shown in the black line.}
    \label{fig:8variaveis}
\end{figure}   

\end{widetext}

The distributions obtained are shown in figure \ref{fig:8variaveis}. The standard deviation for each month is shown in table \ref{tab:std8quantum}, and the monthly mean absolute error for the two experiments is shown in table \ref{tab:mae8quantum}. The cumulative variance of the data contained in this database concerning the original database with nineteen features is 68.69\%.

In the quantum models containing 8 features, experiment 1 with a single variational layer generally showed the best accuracy, with a minimum mean absolute error of 346.10 and a 12-month mean absolute error of 709.76 ± 297.93.

In the second experiment, the best result was again the one in which 5 variational layers were considered, where the lowest mean absolute error obtained was 441.65, and the monthly mean was 980.79 ± 239.93. As occurred in the experiment with 4 features, there is no statistical difference between this result and the result obtained with 3 variational layers. However, the results of this model showed considerably less variation than the results obtained with just one variational layer and also a considerably lower mean. 

In this case, the forecast means were more distributed compared to the target in the best result of experiment 1, but the results of the second experiment echoed the trend of exceeding the actual values, generating a production signal above actual demand.

\subsubsection{19 Features}
In the third considered case, the dataset was tested using all the features for which data was available for 2019. Features that did not have data for 2019 were discarded. The alternative to discarding these features would be to perform inference for 2019 data. However, the amount of data that would be inferred would represent approximately 1/4 of the dataset, a portion that would compromise the model's performance.

In order to train the model with this dataset, no transformation other than data standardization was carried out. However, from the 8th month of the test set onwards, it was identified that one of the features had increased significantly in relation to the others. For this reason, it was decided to maintain this data to observe the effects that this increase in one of the features would have on the results.

\begin{widetext}

\begin{figure}[H]
    \centering
    \includegraphics[scale = 0.5]{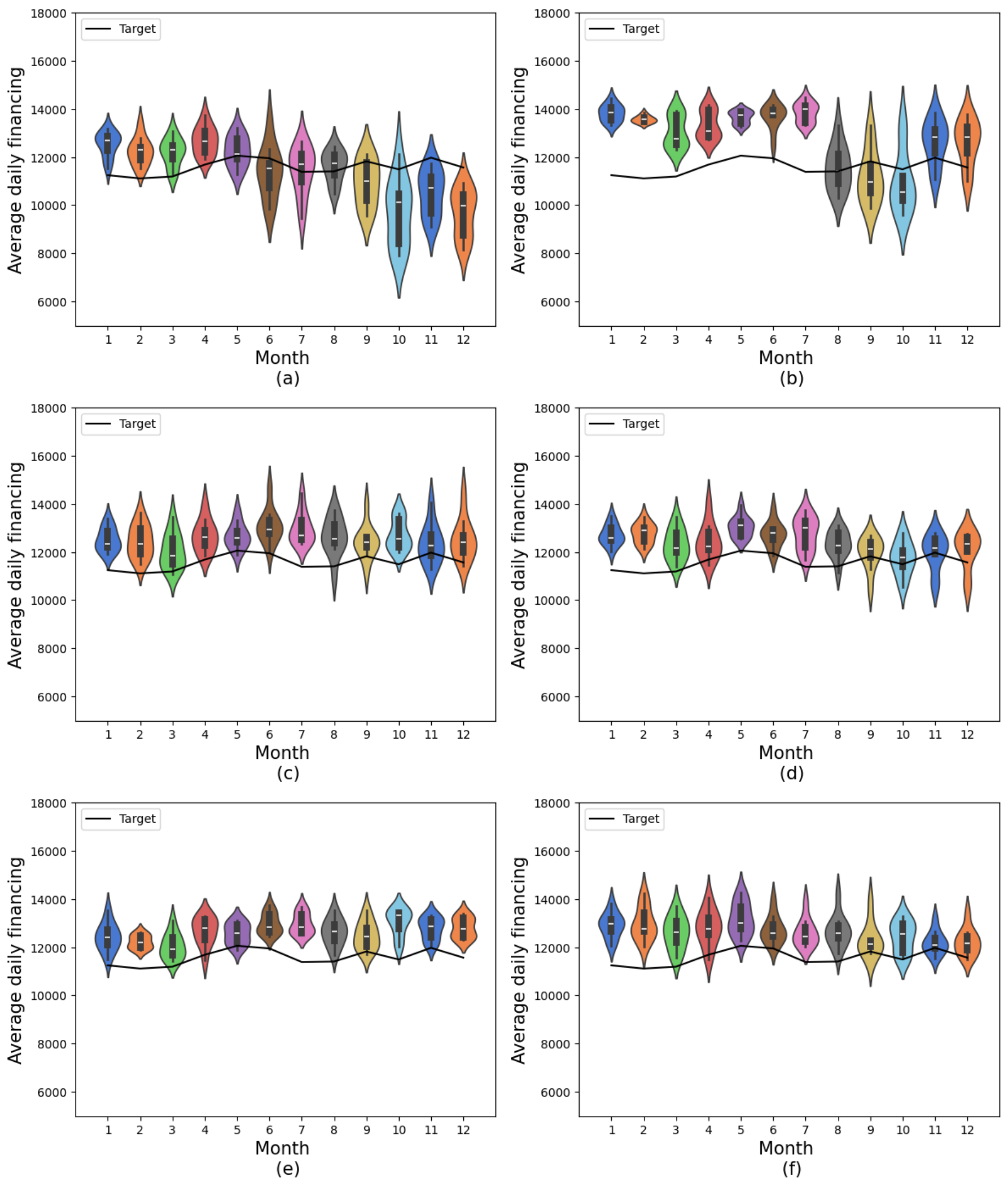}
    \caption{Predictions for quantum models with 19 features. Figure (a) shows the quantum experiment 1 with 1 layer, figure (b) quantum experiment 2 with 1 layer, figure (c) quantum experiment 1 with 3 layers, figure (d) quantum experiment 2 with 3 layers, figure (e) quantum experiment 1 with 5 layers and (f) quantum experiment 2 with 5 layers. The x-axis shows the model's training months, while the y-axis represents average daily financing. The distributions obtained from 10 experiments are shown in the colored violin graphs, while the actual values are shown in the black line.}
    \label{fig:quantico19}
\end{figure}  

\end{widetext}

The distributions obtained are shown in the subfigures provided in figure \ref{fig:quantico19}. The standard deviation for each month is shown in table \ref{tab:std19quantum}, and the monthly and annual mean absolute error for the two experiments are shown in table \ref{tab:mae19quantum}. Since this database contains all the features from the original dataset, the cumulative variance of the data contained in this database is the total variance of the original set, i.e., 100\%.

In experiment 1, which was performed with 19 features, there was no significant difference between the results obtained with 1, 3, and 5 variational layers. For all three cases, the minimum absolute mean errors obtained were 522.95, 638.46, and 533.07, and the 12-month means were 1141.58 ± 465.98, 1054.52 ± 298.52, and 1051.39 ± 311.39 respectively. 
In experiment 2, the smallest mean absolute monthly errors are 758.83, 552.07, and 365.82, respectively, and the monthly means are 1630.98 ± 625.66, 995.71 ± 374.16, and 1074.90 ± 450.09. In this case, however, the first result presents a larger error linked to a larger standard deviation.

\subsection{Classical Experiments} \label{Classical experiments}

\begin{widetext}

\begin{figure}[H]
    \centering
    \includegraphics[scale = 0.45]{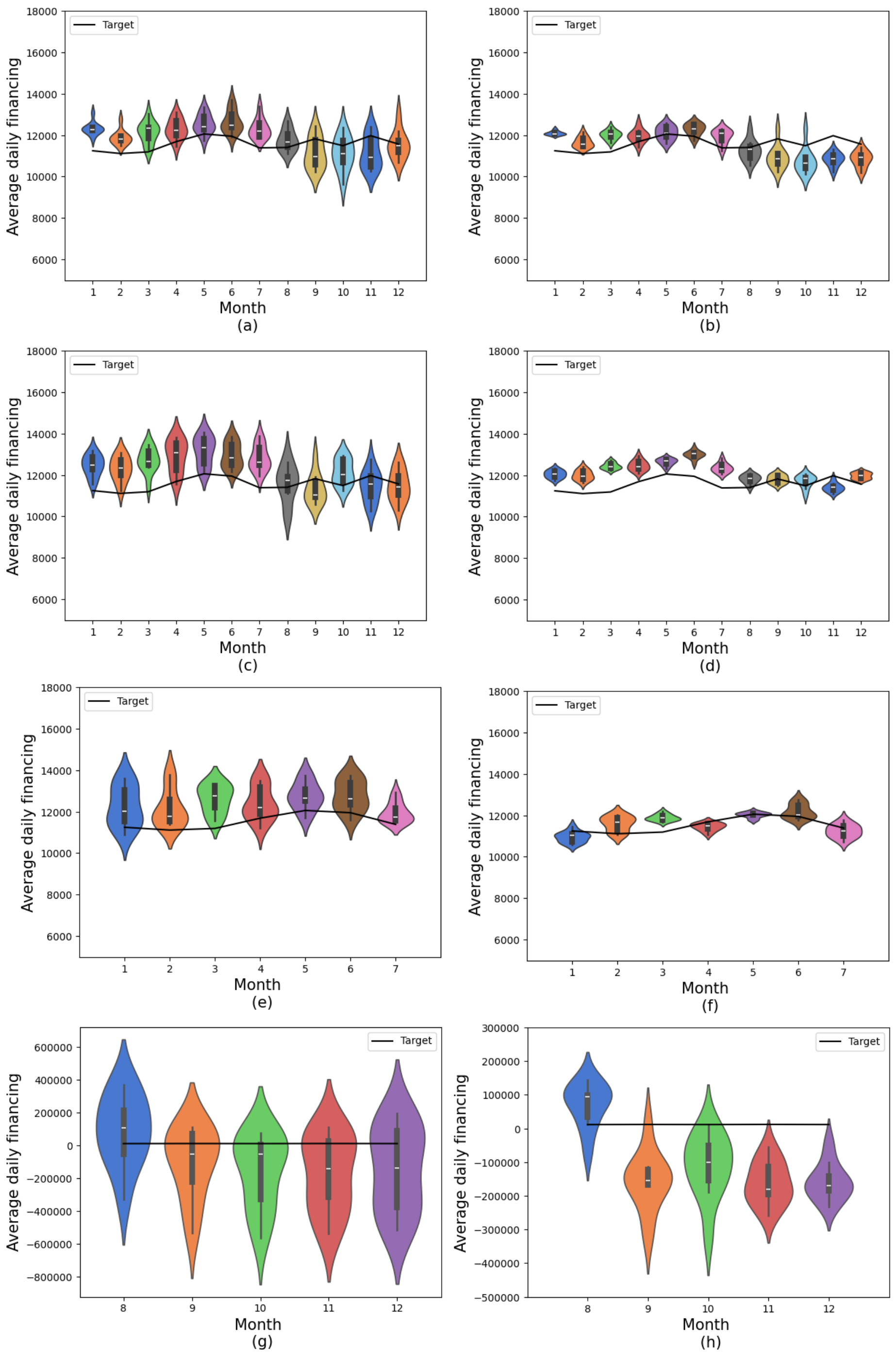}
    \caption{Predictions for classical models. Figure (a) shows the classical experiment 1 with 4 features, figure (b) Classical Experiment 2 with 4 features, figure (c) Classical Experiment 1 with 8 features, figure (d) Classical Experiment 2 with 8 features, figure (e) the first 7 months of classical experiment 1 with 19 features, figures (f) the first 7 months of classical experiment 2 with 19 features, (g) the last 5 months of classical experiment 1 with 19 features and figure (h) the last 5 months of classical experiment 2 with 19 features. The x-axis shows the model's training months, while the y-axis represents average daily financing. The distributions obtained from 10 experiments are shown in the colored violin graphs, while the actual values are shown in the black line.}
    \label{fig:classico19}
\end{figure}  
\end{widetext}

The classical experiments were performed as a way of comparing the results obtained with those of traditionally used classical methods. For this purpose, a classical RNN with 128 and 1024 neurons in the recurrent layer was considered. Table \ref{fig:classico19} shows the results obtained using this model, and convergence graphs are shown in section \ref{Convergence}.

In experiment 1, involving 4 and 8 features, the results were similar, so the 12-month means of the mean absolute errors were 774.93 ± 199.39 and 997.12 ± 331.18, respectively. However, when all 19 initial features were taken into account, these results were skewed by the variable that had significantly higher values in the last five months, so the error increased significantly, and the 12-month mean was 90480.42 ± 111210.54. Given that this increase in error and error variance only occurs in the last few months of training, the results for 19 characteristics are presented in two graphs, as the results scale has changed.

Disregarding the last 5 months, the model showed a mean absolute error of 344.19 with a mean standard deviation of 197.55, which is significantly better than all the other models (classical or quantum) presented.

In experiment 2, involving 4 and 8 features, the results also behaved similarly, so that the 12-month means of the absolute mean errors were 643.12 ± 293.77 and 683.30 ± 332.37. As was the case in the first experiment, when all 19 initial features were considered, these results were biased by the variable that had significantly higher values in the last five months, so that the error increased significantly, and the 12-month mean in this experiment was 7965.14 ± 61675.02. The results of this experiment, considering 19 features, were also presented in two graphs since the scale of the results was modified.

\subsection{Simulation environment} \label{environment}
All the simulations were performed in the PennyLane quantum computing software development kit \cite{pennylane}, developed by the quantum computing company Xanadu \cite{xanadu}, also using the TensorFlow machine learning library \cite{tensorflow}. The simulations were carried out in an HPC environment on Intel Xeon Platinum 8260L processors. The simulations involving 19 features were performed on 17 cores, while the simulations involving 4 and 8 features were performed on a single core. 

The execution times for each sample of the quantum model are shown in table \ref{tab:tempo} and for the classical model in table \ref{tab:tempo_classico}. The total computing time of the runs is given by the values in the table multiplied by ten since ten samples were extracted for each model.

\begin{table}[H]
\centering
\caption{Processing times for quantum models with 4, 8, and 19 features and 1, 3, and 5 layers.}
\begin{tabular}{|cc|c|c|c|ll}
\cline{1-5}
\multicolumn{2}{|l|}{}                              & 1 layer & 3 layers & 5 layers &  &  \\ \cline{1-5}
\multicolumn{1}{|c|}{{4 features}}  & Experiment 1 & 1min30  & 3min   & 4min  &  &  \\ \cline{2-5}
\multicolumn{1}{|c|}{}                              & Experiment 2 & 1min30  & 4min   & 5min  &  &  \\ \cline{1-5}
\multicolumn{1}{|c|}{8 features}  & Experiment 1 & 3min    & 6min   & 10min &  &  \\ \cline{2-5}
\multicolumn{1}{|c|}{}                              & Experiment 2 & 3min    & 6,5min & 10min &  &  \\ \cline{1-5}
\multicolumn{1}{|c|}{{19 features}} & Experiment 1 & 1h      & 1h30   & 3h30  &  &  \\ \cline{2-5}
\multicolumn{1}{|c|}{}                              & Experiment 2 & 1h      & 2h20   & 3h30  &  &  \\ \cline{1-5}
\end{tabular}
\label{tab:tempo}
\end{table}

\begin{table}[H]
    \centering
    \caption{Processing times for the classical models with 4, 8 and 19 features.}
    \begin{tabular}{|c|c|c|}
    \hline
                      & Experiment 1 & Experiment 2\\ \hline
        4 features   & 13min   & 32min30 \\  \hline
        8 features   & 14min30 & 36min \\ \hline
        19 features  & 14min30 & 36min \\ \hline
    \end{tabular}
    \label{tab:tempo_classico}
\end{table}

Given that the demand forecasting problem was solved through a simulation and not on a quantum computer, it is not possible to correlate the processing times obtained with those obtained on a quantum computer. However, the rapid convergence of the models can be interpreted as an indication of the shorter execution times required. 

Processing time is often pointed out as an advantage of quantum computing since some quantum algorithms have an advantage over the best classical algorithms that perform the same task. For instance, Shor's algorithm and Grover's algorithm are able to perform tasks in exponentially and quadratically less time than a classic computer, respectively. However, when analyzing the computational advantages, other metrics must be taken into account, such as the accuracy of the results and the savings in computational and energy resources.

\subsection{Convergence} \label{Convergence}

\begin{figure}[H]
 \includegraphics[scale=0.6]{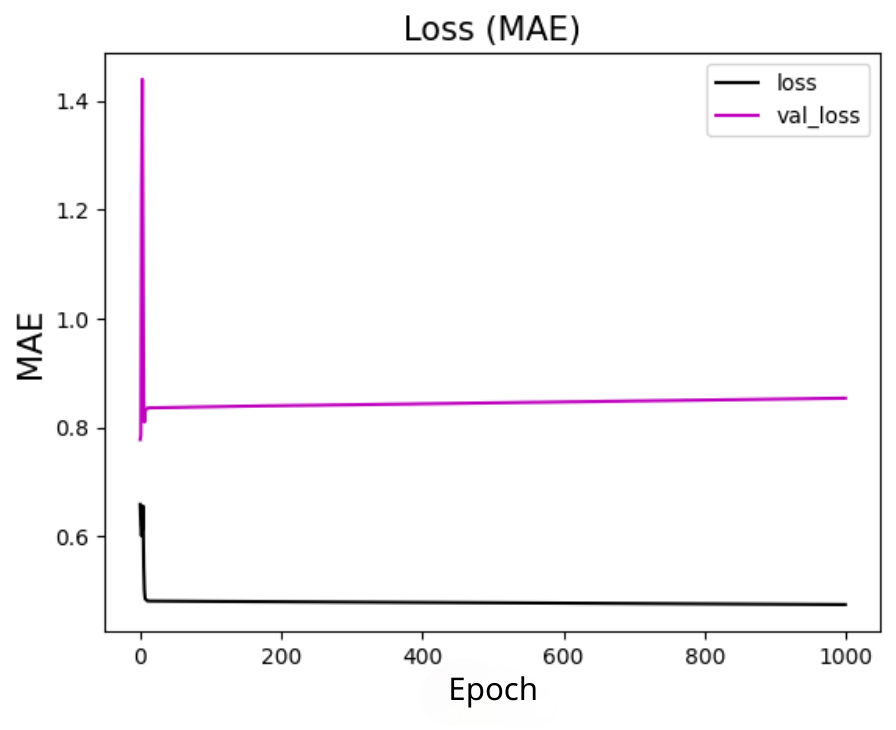}
 \caption{Convergence of the quantum model with the set of 4 features and 1 layer with 1000 epochs. The x-axis shows the training epochs, while the y-axis shows the mean absolute error (standardized values). The black curve shows the test loss, while the magenta curve shows the validation loss.} \label{fig:rapidaconvergenciaa}
\end{figure}

The research on the convergence of quantum models showed that the results converge before the first 30 training epochs, so the 10 experiments used for the statistical analysis of each model were carried out using only 30 epochs. This decision was based on preliminary tests considering training with 1000 epochs, which showed rapid convergence of the models as shown in figure \ref{fig:rapidaconvergenciaa} for 4 features, with marginal or zero improvements in performance from that point onwards. Therefore, this approach optimizes the use of computational resources and avoids over-training. In addition, although the models were simulated in a classical environment, quantum resources are currently scarce, so the predictive quality of the models linked to fast and stable convergence should be considered an advantage of these models. The convergence graphs of the models are shown in figures \ref{fig:loss4features}, \ref{fig:loss8features} and \ref{fig:loss19features}.

\subsubsection{Quantum Models}  \label{Quantum Models}

\begin{widetext}

\begin{figure}[H]
    \centering
    \includegraphics[scale = 0.5]{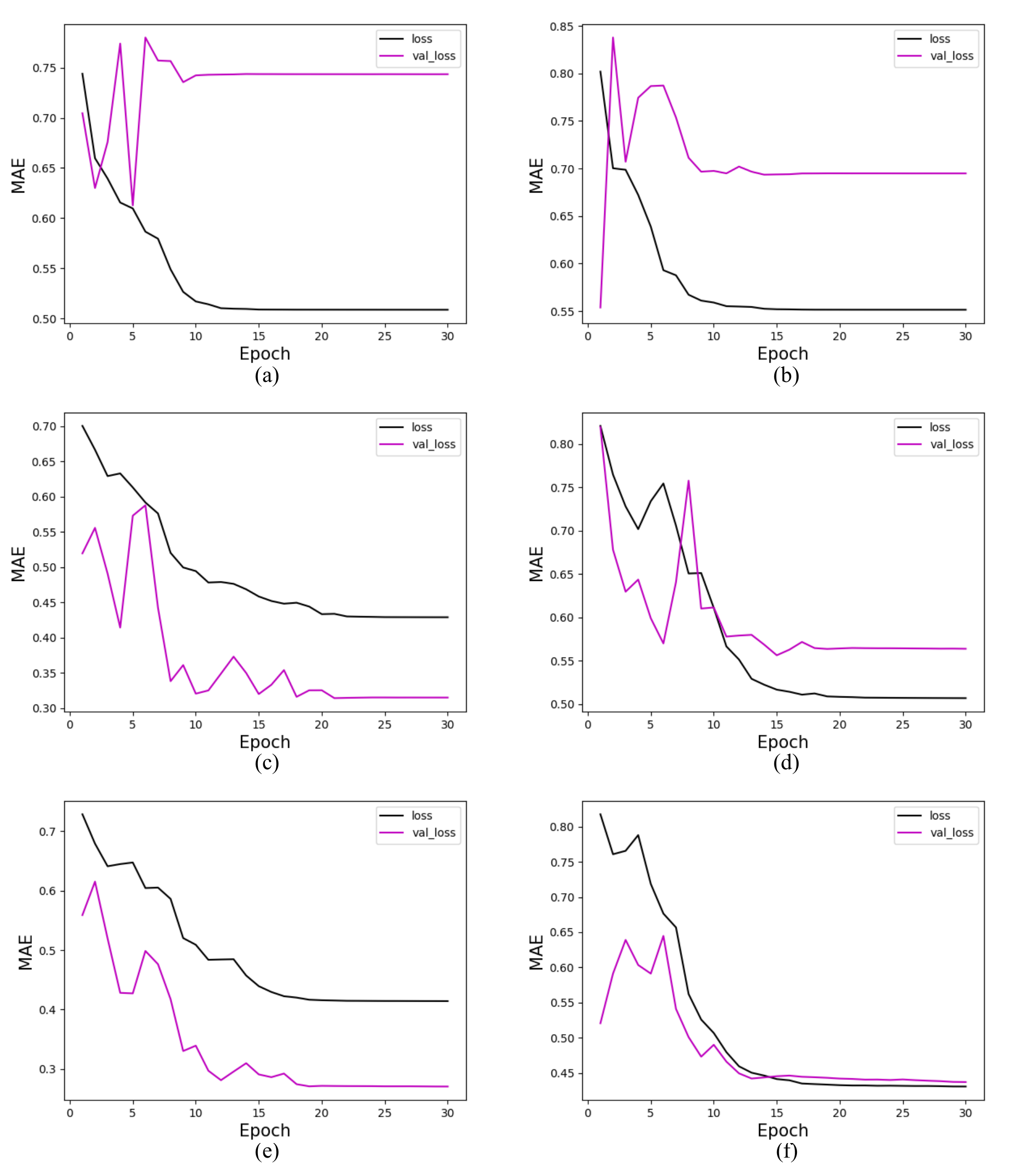}
    \caption{Loss for quantum models with 4 features. Figure (a) shows the loss for quantum experiment 1 with 1 layer, figure (b) for quantum experiment 2 with 1 layer, figure (c) for quantum experiment 1 with 3 layers, figure (d) for quantum experiment 2 with 3 layers, figure (e) for quantum experiment 1 with 5 layers and (f) for quantum experiment 2 with 5 layers. The x-axis shows the training epochs, while the y-axis shows the mean absolute error (standardized values). The black curve shows the test loss, while the magenta curve shows the validation loss.}

    \label{fig:loss4features}
\end{figure}

\begin{figure}[H]
    \centering
    \includegraphics[scale = 0.5]{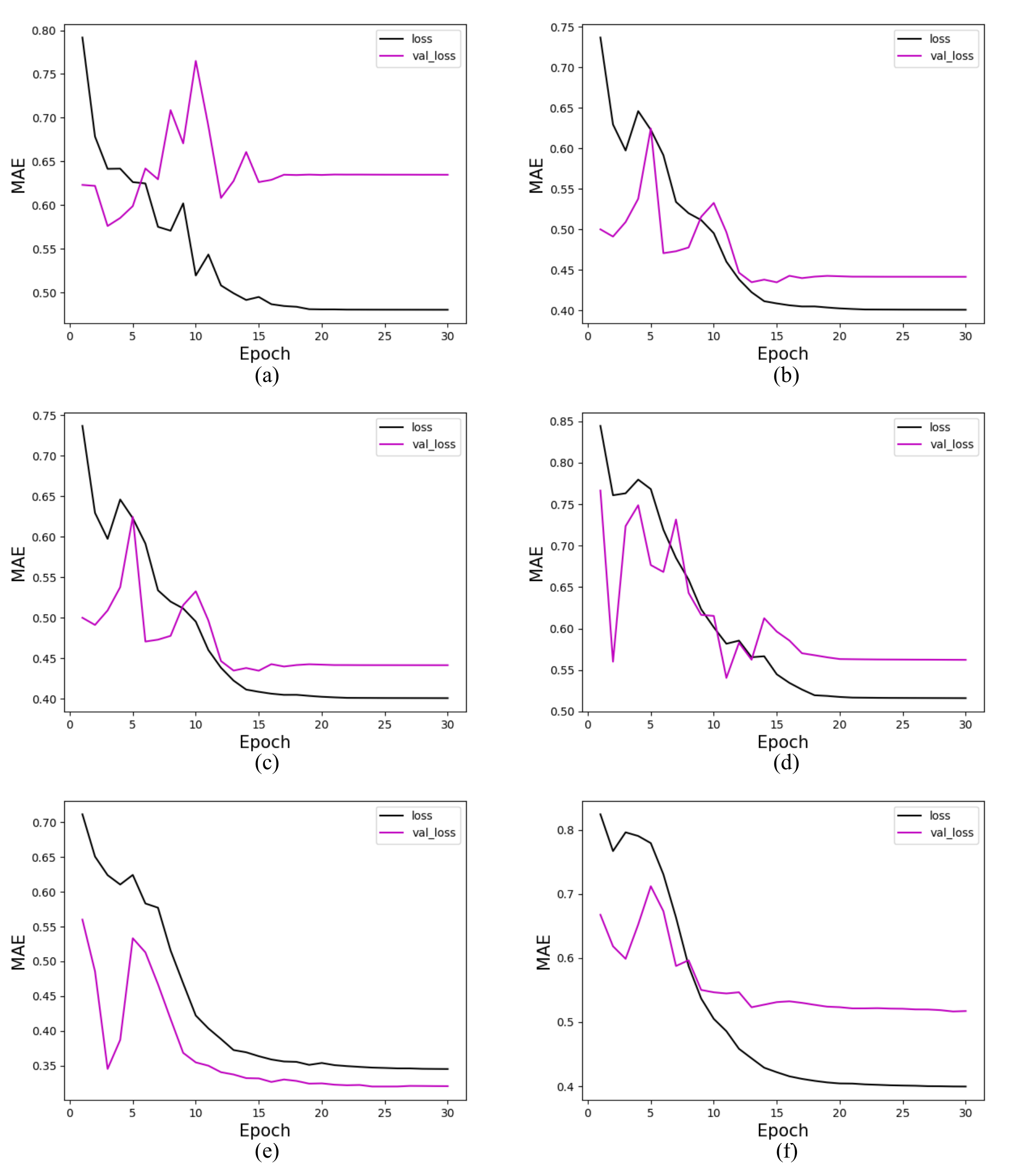}
    \caption{Loss for quantum models with 8 features. Figure (a) shows the loss for quantum experiment 1 with 1 layer, figure (b) for quantum experiment 2 with 1 layer, figure (c) for quantum experiment 1 with 3 layers, figure (d) for quantum experiment 2 with 3 layers, figure (e) for quantum experiment 1 with 5 layers and (f) for quantum experiment 2 with 5 layers. The x-axis shows the training epochs, while the y-axis shows the mean absolute error (standardized values). The black curve shows the test loss, while the magenta curve shows the validation loss.}
    \label{fig:loss8features}
\end{figure}

\begin{figure}[H]
    \centering
    \includegraphics[scale = 0.5]{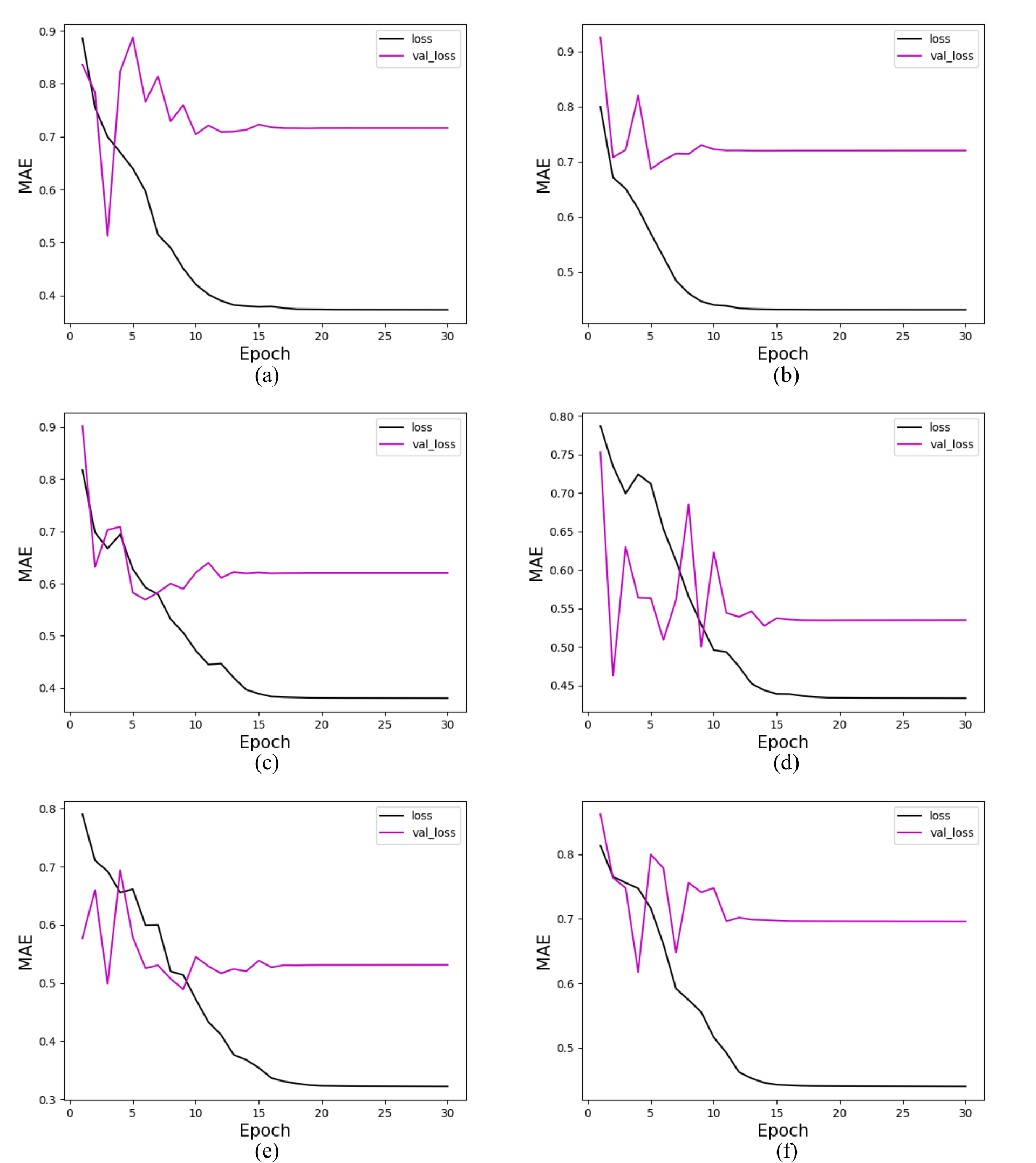}
    \caption{Loss for quantum models with 19 features. Figure (a) shows the loss for quantum experiment 1 with 1 layer, figure (b) for quantum experiment 2 with 1 layer, figure (c) for quantum experiment 1 with 3 layers, figure (d) for quantum experiment 2 with 3 layers, figure (e) for quantum experiment 1 with 5 layers and (f) for quantum experiment 2 with 5 layers. The x-axis shows the training epochs, while the y-axis shows the mean absolute error (standardized values). The black curve shows the test loss, while the magenta curve shows the validation loss.}
    \label{fig:loss19features}
\end{figure}

\end{widetext}

\subsubsection{Classical Models}  \label{Classical Models}

\begin{widetext}

\begin{figure}[H]
    \centering
    \includegraphics[scale = 0.5]{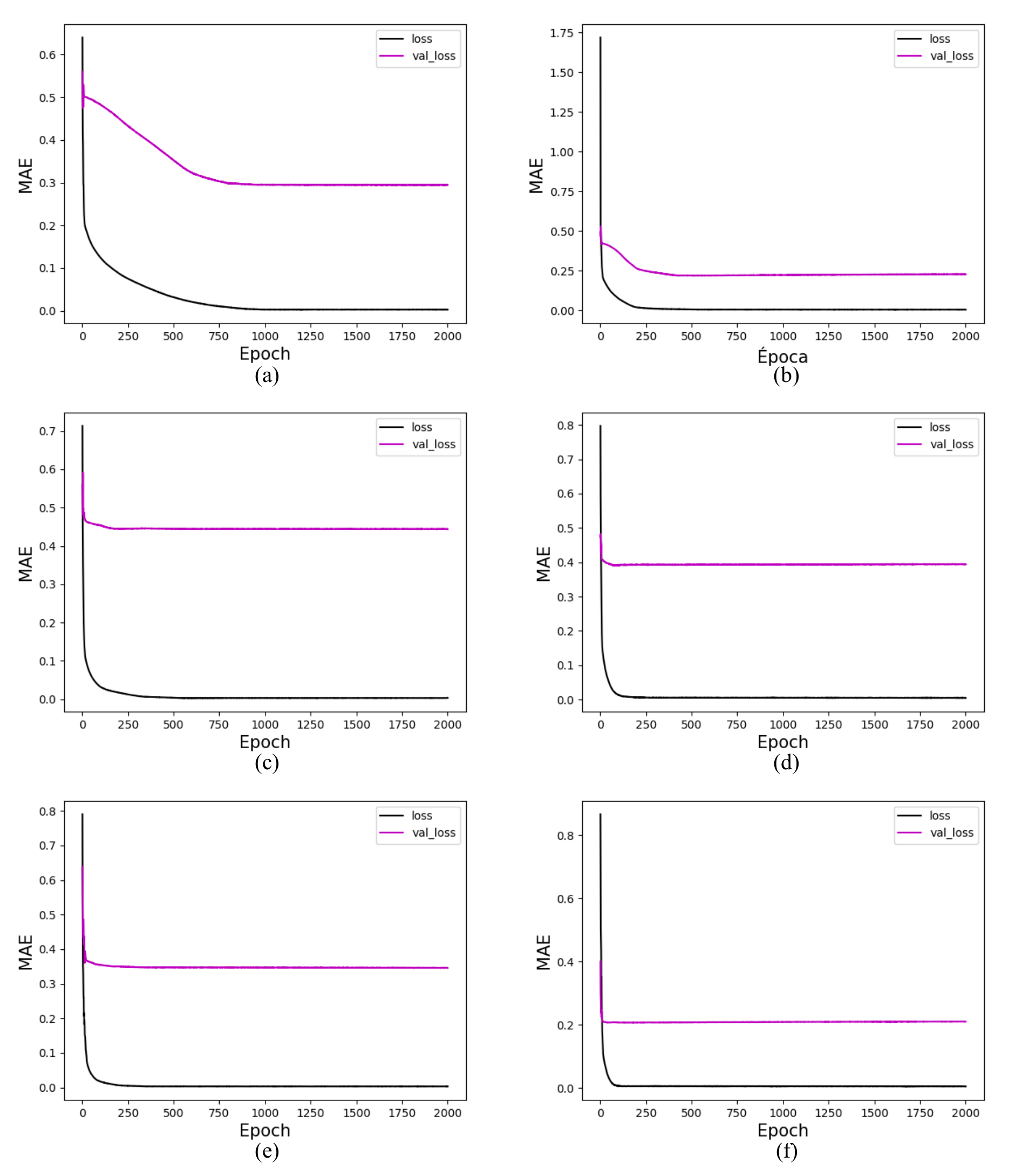}
    \caption{Loss for classical models. Figure (a) shows the loss for classical experiment 1 with 4 features, figure (b) for classical experiment 2 with 4 features, figure (c) for classical experiment 1 with 8 features, figure (d) for classical experiment 2 with 8 features, figure (e) for quantum experiment 1 with 19 features and (f) for classical experiment 2 with 19 features. The x-axis shows the training epochs, while the y-axis shows the mean absolute error (standardized values). The black curve shows the test loss, while the magenta curve shows the validation loss.}   
    \label{fig:lossclassica}
\end{figure}

\end{widetext}

\section{Conclusion}
\label{sec-conclusions}

Here, a quantum neural network was utilized for the first time to solve the problem of predicting short-term demand for used vehicles. {The tests were carried out on datasets with 4, 8, and 19 features. The results were compared with those obtained using a classical recurrent neural network, showing similarities of the models in terms of accuracy in the best case for each one, but with the quantum model using fewer features and parameters and converging in fewer epochs than the classical model. In addition, the quantum model showed less bias towards problematic features in the scenarios with the largest number of features considered. Thus, the results show evidence that quantum models can be excellent candidates for future implementation of this task in large-scale quantum computers. These results can possibly be extended to other predictions of interest to the financial sector, creating a new way of forecasting in the financial industry.} These results could be better explored in subsequent stages by testing other quantum models, including quantum analogs to classical recurrent neural networks, and comparing the results with more robust variations of classical recurrent neural networks.

\section{ACKNOWLEDGEMENTS}

The authors would like to thank the Brazilian bank BV for providing the data, technical and financial support. They also thank SENAI CIMATEC Supercomputing Center for Industrial Innovation for infrastructure access and technical support. 

C. Cruz and M.H.F. da Silva thank the Bahia State Research Support Foundation (FAPESB) for financial support (grant numbers APP0041/2023 and PPP0006/2024). This work has been partially supported by QuIIN - EMBRAPII CIMATEC Competence Center in Quantum Technologies, with financial resources from the PPI IoT/Manufatura 4.0 of the MCTI grant number 053/2023, signed with EMBRAPII.

\bibliographystyle{unsrt}
\bibliography{references}

\appendix

\section{Standard Deviation} 

The monthly standard deviation obtained in each experiment is presented in this section. The subsection \ref{subsection:quantumstd} shows the deviations obtained in the quantum experiments, while the subsection \ref{subsection:classicalstd} shows the deviations obtained in the classical experiments.

\begin{widetext}
\subsection{Quantum Experiments} \label{subsection:quantumstd}

The tables \ref{tab:std4quantum}, \ref{tab:std8quantum}, and \ref{tab:std19quantum} show the standard deviation obtained in the quantum experiments carried out with 4, 8, and 19 features, respectively. The columns of the tables represent each experiment and the number of layers used, while the rows represent the deviations in each month. The last two rows represent the mean and median deviations over the 12-month period.

\begin{table}[H]
\centering
\begin{tabular}{|c|c|c|c|c|c|c|c|}
\hline
& \multicolumn{3}{|c|}{Experiment 1} & \multicolumn{3}{|c|}{Experiment 2} \\ \hline
& 1 layer & 3 layers & 5 layers & 1 layer & 3 layers & 5 layers \\ \hline
Month 1  & 175.8777 & 449.1271 & 579.0006 & 292.4347 & 781.3115 & 945.6261
\\ \hline
Month 2  & 160.8550 & 454.754 & 541.9076 & 311.3989 & 697.3065 & 928.2651
\\ \hline
Month 3  & 243.3442 & 517.2781 & 590.0444 & 324.5650 & 812.6765 & 100.6944
 \\ \hline
Month 4  & 348.6137 & 574.0108 & 598.8600 & 410.6001 & 871.5712 & 1111.7944
\\ \hline
Month 5  & 384.9921 & 710.9180 & 472.9006 & 355.1420 & 780.3099 & 795.7520
\\ \hline
Month 6  & 443.4344 & 770.3615 & 422.4470 & 326.7457 & 730.8906& 498.5881
\\ \hline
Month 7  & 345.1314 & 893.0815 & 746.0978 & 629.6234 & 858.7378 & 506.8464
\\ \hline
Month 8  & 815.3586 & 990.5417 & 737.8833 & 707.0781 & 1085.2201 & 506.1400
\\ \hline
Month 9  & 543.9914 & 982.4916 & 804.3594 & 871.3573 & 1055.3599 & 458.0408
\\ \hline
Month 10 & 643.1245 & 810.9172& 597.7432 &578.6797 &931.9728 & 502.1143
\\ \hline
Month 11 & 738.7717 & 658.1062& 395.4243 &347.0588 &1020.1457 & 794.2041
\\ \hline
Month 12 & 905.4925 & 695.5312& 451.0824 &302.2530 & 1077.0132 & 679.5169 \\ \hline
Mean & 479.0823 & 708.8950 & 578.1459 & 454.7447 & 891.8763 & 727.3819\\ \hline
Median & 414.2133 & 703.2246 & 584.5225 & 351.1004 & 865.1545 & 736.8605\\ \hline
\end{tabular}
\caption{Monthly standard deviation for quantum experiments with 4 features. The columns represent each experiment with a number of 1, 3, and 5 variational layers, and the lines represent the months. The last two lines show the 12 month mean and median of the mean absolute error, respectively.}
\label{tab:std4quantum}
\end{table}

\begin{table}[H]
    
    \centering
    \begin{tabular}{|c|c|c|c|c|c|c|c|}
    \hline
    & \multicolumn{3}{|c|}{Experiment 1} & \multicolumn{3}{|c|}{Experiment 2} \\ \hline
       & 1 layer & 3 layers & 5 layers & 1 layer & 3 layers & 5 layers \\ \hline
Month 1  & 160.8833  & 252.3982 & 326.4278 & 121.1870 & 540.7551 & 466.1218 \\ \hline
Month 2  & 389.2180  & 321.1851 & 321.9523 & 115.6250 & 544.9329 & 578.9061 \\ \hline
Month 3  & 199.1200  & 407.9355 & 427.0799 & 116.9056 & 628.8543 & 312.7844 \\ \hline
Month 4  & 317.9407  & 397.5779 & 462.1943 & 276.6656 & 595.7785 & 338.2273 \\ \hline
Month 5  & 406.4636  & 597.9564 & 480.5169 & 345.8178 & 437.5681 & 255.3891 \\ \hline
Month 6  & 710.3341  & 482.3563 & 550.8562 & 235.6304 & 739.2531 & 322.0360 \\ \hline
Month 7  & 557.0957  & 722.5253 & 642.4665 & 575.8245 & 602.2518 & 353.3172 \\ \hline
Month 8  & 1059.4827 & 621.9092 & 885.9215 & 693.9122 & 616.6715 & 352.7128 \\ \hline
Month 9  & 758.2744  & 793.3121 & 874.5666 & 801.1654 & 976.0197 & 579.2410 \\ \hline
Month 10 & 841.0811  & 915.8552 & 820.2909 & 697.0682 & 655.6456 & 514.6416\\ \hline
Month 11 & 793.1211  & 887.0502 & 612.1734 & 269.6858 & 615.5451 & 507.4711\\ \hline
Month 12 & 1245.5292 & 529.3304 & 646.7094 & 483.5458 & 799.0230 & 362.0224 \\ \hline
Mean & 619.8787 & 577.4493 & 587.5963 & 394.4194 & 646.0249 & 411.9059\\ \hline
Median & 633.7149 & 563.6434 & 581.5148 & 311.2417
 & 616.1083 & 357.6698\\ \hline
    \end{tabular}
    \caption{Monthly standard deviation for quantum experiments with 8 features. The columns represent each experiment with a number of 1, 3, and 5 variational layers, and the lines represent the months. The last two lines show the 12 month mean and median of the mean absolute error, respectively.}    
    \label{tab:std8quantum}
\end{table}

\begin{table}[H]
    \centering
    \begin{tabular}{|c|c|c|c|c|c|c|c|}
    \hline
    & \multicolumn{3}{|c|}{Experiment 1} & \multicolumn{3}{|c|}{Experiment 2} \\ \hline
       & 1 layer & 3 layers & 5 layers & 1 layer & 3 layers & 5 layers \\ \hline
Month 1  & 516.7164  & 494.9935 & 578.7277 & 355.9007  & 427.1650 & 521.0726 \\ \hline
Month 2  & 572.1335  & 685.2427 & 280.9679 & 145.9410  & 421.7772 & 693.0295 \\ \hline
Month 3  & 586.1252  & 738.5064 & 532.6228 & 675.5994  & 658.4324 & 688.9236 \\ \hline
Month 4  & 598.8343  & 669.0032 & 586.9466 & 590.2582  & 758.9672 & 748.5518 \\ \hline
Month 5  & 639.3816  & 539.5154 & 444.1018 & 266.6220  & 437.0098 & 685.1440 \\ \hline
Month 6  & 1091.7456 & 705.1529 & 436.6959 & 519.9328  & 534.6220 & 579.4321 \\ \hline
Month 7  & 999.3549  & 665.8485 & 439.2488 & 413.0658  & 689.7512 & 570.5552 \\ \hline
Month 8  & 647.7254  & 797.2639 & 559.3305 & 932.4816  & 570.7285 & 670.7190 \\ \hline
Month 9  & 967.2389  & 707.2880 & 587.0741 & 1128.5844 & 684.2530 & 737.8799 \\ \hline
Month 10 & 1389.6125 & 641.6429 & 565.2284 & 1301.0933 & 705.1833 & 659.6317 \\ \hline
Month 11 & 953.2446  & 811.4675 & 466.9461 & 920.2317  & 748.9888 & 479.6019 \\ \hline
Month 12 & 995.1870  & 863.9411 & 424.4394 & 959.6248  & 798.4880 & 565.5388 \\ \hline
Mean & 829.7750  & 693.3222 & 491.8608 & 648.1113 & 619.6139 & 633.3400 \\ \hline
Median & 800.4850 & 695.1979 & 499.7845 & 632.9288 & 671.3427 & 665.1754\\ \hline
    \end{tabular}
    \caption{Monthly standard deviation for quantum experiments with 19 features. The columns represent each experiment with a number of 1, 3, and 5 variational layers, and the lines represent the months. The last two lines show the 12 month mean and median of the mean absolute error, respectively.}
    \label{tab:std19quantum}
\end{table}
\end{widetext}

\subsection{Classical Experiments}\label{subsection:classicalstd}

Table \ref{tab:classicalstd} shows the standard deviation obtained in the classical experiments carried out with 4, 8, and 19 features. The columns of the table represent each experiment and the number of features used, while the rows represent the deviations in each month. The last two rows represent the average and median deviations over the 12-month period.

\begin{widetext}

\begin{table}[H]
    \centering
    \begin{tabular}{|c|c|c|c|c|c|c|}
    \hline
    & \multicolumn{3}{|c|}{Classical experiment 1} & \multicolumn{3}{|c|}{Classical experiment 2}\\ \hline
    & 4 features & 8 features & 19 features & 4 features & 8 features & 19 features \\ \hline
Month 1  &  308.17  &  515.30   & 983.76 &  99.29   & 189.26   &  263.49   \\ \hline
Month 2  &  354.86  &  592.63   & 928.49 &  258.99  & 240.52   &  368.53   \\ \hline
Month 3  &  528.00  &  597.08   & 684.88 &  221.78  & 155.91   &  158.13   \\ \hline
Month 4  &  495.66  &  797.64   & 817.02 &  321.39  & 252.95   &  190.64   \\ \hline
Month 5  &  472.31  &  703.64   & 676.05 &  333.65  & 171.95   &  136.29    \\ \hline
Month 6  &  552.07  &  611.75   & 748.18 &  287.49  & 197.82   &  353.89   \\ \hline
Month 7  &  484.05  &  613.97   & 494.19 &  346.032 & 230.98   &  327.44   \\ \hline
Month 8  &  521.11  &  929.49   & 218484.99 &  492.37  & 213.54   &  66181.53   \\ \hline
Month 9  &  766.99  &  729.89   & 217021.29  &  573.75  & 204.83   &  87269.59    \\ \hline
Month 10 &  809.57  &  621.17   & 224172.77 & 623.91  & 312.91   &  94615.78   \\ \hline
Month 11 &  797.50  &  760.97   & 231876.29 &  322.52  & 218.21   &  63811.92    \\ \hline
Month 12 &  697.86  &  751.18   & 260196.92 &  385.86  & 154.92   &  55942.54    \\ \hline
Mean  & 565.68 & 685.39 & 96423.73 & 355.59   & 211598   &  30801.65 \\ \hline
Median  & 524.55 & 662.40 & 956.13 & 328.09 & 209.19   &  361.21 \\ \hline
    \end{tabular}
    \caption{Monthly standard deviation for classical experiments with 4, 8, and 19 features. The columns represent each experiment with a number of 4, 8, and 19 features, and the lines represent the months. The last two lines show the 12 month mean and median of the mean absolute error, respectively.}
    \label{tab:classicalstd}
\end{table}

\end{widetext}
\section{Mean Absolute Error}

The mean absolute error obtained in each experiment is presented in this section. The subsection \ref{subsection:quantummae} presents the errors obtained in the quantum experiments, while the subsection \ref{subsection:classicalmae} presents the errors obtained in the classical experiments.

\subsection{Quantum Experiments}\label{subsection:quantummae}

The tables \ref{tab:mae4quantum}, \ref{tab:mae8quantum}, and \ref{tab:mae19quantum} show the mean absolute error obtained in the quantum experiments carried out with 4, 8, and 19 features, respectively. The columns of the tables represent each experiment and the number of layers used, while the rows represent the errors in each month. The last row represents the average of the mean absolute errors over the 12-month period.

\begin{widetext}

\begin{table}[H]
    \centering
    \begin{tabular}{|c|c|c|c|c|c|c|}
    \hline
    & \multicolumn{3}{|c|}{Experiment 1} & \multicolumn{3}{|c|}{Experiment 2} \\ \hline
       & 1 layer & 3 layers & 5 layers & 1 layer & 3 layers & 5 layers \\ \hline
Month 1  & 650.18 & 401.92 & 530.33 & 792.33 & 1463.55 & 871.18     \\ \hline
Month 2  & 1175.84 & 536.48 & 646.12 & 805.95 & 1469.63 & 951.18    \\ \hline
Month 3  & 635.44 & 485.10 & 496.18 & 1117.42 & 1479.38 & 944.70    \\ \hline
Month 4  & 299.90 & 453.46 & 513.00 & 804.10 & 1120.72 & 929.66     \\ \hline
Month 5  & 334.44 & 707.43 & 858.72 & 1560.78 & 738.46 & 673.11     \\ \hline
Month 6  & 493.43 & 658.21 & 842.96 & 2249.48 & 698.33 & 411.97     \\ \hline
Month 7  & 733.64 & 886.78 & 717.68 & 2770.17 & 912.10 & 626.01     \\ \hline
Month 8  & 1144.93 & 1027.13 & 827.18 & 2888.23 & 1013.24 & 867.35  \\ \hline
Month 9  & 566.27 & 1390.83 & 1297.83  & 3154.11 & 885.93 & 522.42  \\ \hline
Month 10  & 577.18 & 898.56 & 698.99 & 2561.83 & 1130.34 & 742.62   \\ \hline
Month 11  & 596.82 & 1036.78 & 1036.90 &  1979.68 & 821.36 & 849.04 \\ \hline
Month 12  & 982.13 & 792.25 & 641.45 & 2048.10 & 1066.48 & 1042.62  \\ \hline
Mean  & 682.52$\pm$284.14 & 772.91$\pm$292.49 & 758.94$\pm$234.12 &  1894.35$\pm$865.87 & 1066.63$\pm$279.64 & 785.98$\pm$192.21 \\ \hline
    \end{tabular}
    \caption{Monthly mean absolute error for quantum experiments with 4 features. The columns represent each experiment with a number of 1, 3, and 5 variational layers, and the lines represent the months. The last line shows the 12-month mean of the mean absolute error.}
    \label{tab:mae4quantum}
\end{table}

\begin{table}[H]
    \centering
    \begin{tabular}{|c|c|c|c|c|c|c|}
    \hline
    & \multicolumn{3}{|c|}{Experiment 1} & \multicolumn{3}{|c|}{Experiment 2} \\ \hline
       & 1 layer & 3 layers & 5 layers & 1 layer & 3 layers & 5 layers \\ \hline
Month 1  & 512.12 & 732.27 & 401.66 & 1097.63 & 1217.56 & 951.43   \\ \hline
Month 2  & 616.63 & 794.92 & 530.46 & 1091.54 & 1366.40 & 1290.20   \\ \hline
Month 3  & 410.03 & 795.86 & 393.68 & 1316.14 & 1277.38 & 1179.36   \\ \hline
Month 4  & 346.10 & 439.18 & 431.12 & 724.89 & 835.73 & 683.89   \\ \hline
Month 5  & 398.91 & 585.06 & 832.43 & 1327.13 & 647.49 & 441.65   \\ \hline
Month 6  & 850.79 & 427.59 & 632.40 & 2452.52 & 1028.89 & 1035.87   \\ \hline
Month 7  & 760.58 & 767.84 & 921.27 & 2418.72 & 1075.02 & 993.72   \\ \hline
Month 8  & 1008.91 & 782.52 & 1008.66 & 2852.20 & 938.75 & 1180.06   \\ \hline
Month 9  & 705.20 & 1388.40 & 1597.51 & 2948.62 & 884.34 & 941.42   \\ \hline
Month 10  & 692.89 & 952.77 & 1075.67 & 1764.08 & 916.54 & 935.58  \\ \hline
Month 11  & 806.42 & 1036.14 & 1186.70 & 1923.66 & 684.19 & 899.97  \\ \hline
Month 12  & 1408.56 & 596.32 & 881.29 & 1709.09 & 1053.04 & 1236.42  \\ \hline
Mean  & 709.76$\pm$297.93 & 774.90$\pm$266.34 &  824.40$\pm$366.91 & 1802.19$\pm$730.42 & 993.78$\pm$222.06 & 980.79$\pm$239.93 \\ \hline
    \end{tabular}
    \caption{Monthly mean absolute error for quantum experiments with 8 features. The columns represent each experiment with a number of 1, 3, and 5 variational layers, and the lines represent the months. The last line shows the 12-month mean of the mean absolute error.}
    \label{tab:mae8quantum}
\end{table}

\begin{table}[H]
    \centering
    \begin{tabular}{|c|c|c|c|c|c|c|}
    \hline
    & \multicolumn{3}{|c|}{Experiment 1} & \multicolumn{3}{|c|}{Experiment 2} \\ \hline
       & 1 layer & 3 layers & 5 layers & 1 layer & 3 layers & 5 layers \\ \hline
Month 1   & 1290.76 & 1271.09 & 1184.78 & 2580.36 & 1465.18 & 1669.35 \\ \hline
Month 2   & 1138.65 & 1341.46 & 1113.86 & 2454.71 & 1664.07 & 1878.59 \\ \hline
Month 3   & 1012.17 & 864.44 & 852.39 & 1860.12 & 1122.22 & 1412.05   \\ \hline
Month 4   & 990.79 & 990.92 & 1030.96 & 1660.29 & 801.17 & 1150.94    \\ \hline
Month 5   & 522.95 & 651.76 & 533.07 & 1597.31 & 987.06 & 1119.59     \\ \hline
Month 6   & 935.99 & 1132.94 & 1024.33 & 1732.11 & 785.72 & 775.02    \\ \hline
Month 7   & 835.87 & 1570.97 & 1529.91 & 2496.80 & 1508.08 & 1190.74  \\ \hline
Month 8   & 525.35 & 1240.00 & 1177.67 & 758.83  & 967.37 & 1232.49   \\ \hline
Month 9   & 1090.46 & 710.54 & 629.81  & 1014.02 & 592.40 & 564.18    \\ \hline
Month 10  & 2013.58 & 1264.20 & 1509.81 & 1208.45 & 623.42 & 892.39   \\ \hline
Month 11  & 1442.46 & 638.46 & 791.39 & 967.58  & 552.07 & 365.82     \\ \hline
Month 12  & 1899.93 & 977.43 & 1238.75 & 1241.23 & 879.85 & 647.64    \\ \hline
Mean   & 1141.58$\pm$465.98 & 1054.52$\pm$298.52 & 1051.39$\pm$311.39 & 1630.98$\pm$625.66& 995.71$\pm$374.16 &  1074.90$\pm$450.09 \\ \hline
    \end{tabular}
    \caption{Monthly mean absolute error for quantum experiments with 19 features. The columns represent each experiment with a number of 1, 3, and 5 variational layers, and the lines represent the months. The last line shows the 12-month mean of the mean absolute error.}
    \label{tab:mae19quantum}

\end{table}
\end{widetext}

\begin{widetext} 
\subsection{Classical Experiments}\label{subsection:classicalmae}

Table \ref{tab:classicalmae} shows the mean absolute error obtained in the classical experiments carried out with 4, 8, and 19 features. The columns of the table represent each experiment and the number of features used, while the rows represent the error obtained in each month. The last two rows represent the average and deviations over the 12-month period.

\begin{table}[H]
    \centering
    \begin{tabular}{|c|c|c|c|c|c|c|}
    \hline
    & \multicolumn{3}{|c|}{Classical experiment 1} & \multicolumn{3}{|c|}{Classical experiment 2} \\ \hline
    & 4 features & 8 features & 19 features  & 4 features & 8 features & 19 features \\ \hline
Month 1  & 1080.65  & 1214.68 & 1083.70 & 852.70  & 806.36 & 331.93  \\ \hline
Month 2  & 772.14   & 1195.05 & 1083.74 & 549.44  & 886.17 & 497.79  \\ \hline
Month 3  & 934.07   & 1513.97 & 1422.53 & 821.23  & 1255.03 & 700.34 \\ \hline
Month 4  & 658.06   & 1244.07 & 806.78  & 332.87  & 781.47 & 229.77    \\ \hline
Month 5  & 549.08   & 1104.81 & 796.94  & 272.64  & 584.70 & 97.59   \\ \hline
Month 6  & 725.33   & 1006.55 & 894.82  & 380.18  & 1035.19 & 289.33   \\ \hline
Month 7  & 909.29   & 1425.20 & 535.88  & 539.62  & 1021.39 & 262.57   \\ \hline
Month 8  & 477.09   & 663.75 &  191450.99 & 286.32  & 357.82 & 83613.94  \\ \hline
Month 9  & 967.82   & 663.77 & 207996.69 & 1005.15  & 199.17 & 178754.93   \\ \hline
Month 10 & 802.83   & 645.03 & 232189.14 & 850.08  & 302.93 &  129657.59   \\ \hline
Month 11 & 933.40   & 673.35 & 211070.68 & 1136.52  & 557.77 &  175224.64  \\ \hline
Month 12 & 489.41   & 615.22 & 236433.09 & 726.71  & 411.61 &  170439.79  \\ \hline
Mean  & 774.93$\pm$199.39 & 997.12$\pm$331.18 & 90480.42$\pm$111210.54 & 646.12$\pm$293.77 & 683.30$\pm$332.37 & 79695.14$\pm$61675.02 \\ \hline
    \end{tabular}
    \caption{Monthly mean absolute error for classical experiments with 4, 8, and 19 features. The columns represent each experiment with a number of 4, 8, and 19 features, and the lines represent the months. The last line shows the 12-month mean of the mean absolute error.}
    \label{tab:classicalmae}
\end{table}
\end{widetext}

\end{document}